\documentclass[aps,amssymb,includegraphics]{revtex4}                  
\usepackage{graphicx}
\usepackage{amsmath}
\usepackage{float}
\usepackage{subfigure}
\usepackage{mathptmx}
\usepackage{enumerate}

\usepackage{natbib}

\usepackage{pdflscape}
\usepackage{afterpage}
\usepackage{capt-of}
\usepackage{rotating}

\begin{document}
\title{Statistical Physics Models of Belief Dynamics: Theory and Empirical Tests}

\author{Mirta~Galesic}
\email{galesic@santafe.edu}
\affiliation{Santa Fe Institute,
1399 Hyde Park Road, Santa Fe, NM 87501 USA}

\author{D.L.~Stein} 
\email{daniel.stein@nyu.edu}
\affiliation{Department of Physics and Courant Institute of Mathematical Sciences,
New York University, New York, NY 10012 USA and NYU-ECNU Institutes of Physics and Mathematical Sciences at NYU Shanghai, 3663 Zhongshan Road North, Shanghai, 200062, China}

\begin{abstract} 
We build simple computational models of belief dynamics within the framework of discrete-spin statistical physics models, and explore how suitable they are for understanding and predicting real-world belief change on both the individual and group levels. We find that accurate modeling of real-world patterns requires attending to social interaction rules that people use, network structures in which they are embedded, distributions of initial beliefs and intrinsic preferences, and the relative importance of social information and intrinsic preferences. We demonstrate that these model parameters can be constrained by empirical measurement, and the resulting models can be used to investigate the mechanisms underlying belief dynamics in actual societies. We use data from two longitudinal studies of belief change, one on 80~individuals living in an MIT dorm during the 2008 presidential election season, and another on 94~participants recruited from Mechanical Turk during the 2016 presidential election primary season. We find that simple statistical physics-based models contain predictive value for real-world belief dynamics and enable empirical tests of different assumptions about the underlying network structure and the social interaction rules.

\end{abstract}

\maketitle

\section{Introduction}
\label{sec:intro}

One of the most active areas of application of statistical physics outside the physical sciences is the development of models of belief dynamics in social networks~\cite{CFL09,Galam12,SC14,CS73,HL75,CG86,FK96,Redner01,DNAW01,WDA05,BR05}. Although these models have uncovered a wide range of interesting behaviors, many of which are suggestive of phenomena observed in the real world, skepticism (in both the physics and social science communities) remains as to whether such models can be useful in understanding, and predicting social phenomena. While some studies have compared results of these models to observed social phenomena~\cite{CostaFilho99,CostaFilho03,Gonzales04,Situngkir04,BRG16}, this literature remains relatively small and limited to group-level and indirect comparisons with empirical data~\cite{FMR13}. 

Here we aim to directly compare results obtained from a class of simple discrete-spin models of belief dynamics with empirical data obtained through longitudinal surveys of small populations on a variety of political, social, and health issues. We relate parameters of simple Ising and Potts models to empirically measurable cognitive and social factors described in social psychology, animal learning, evolutionary anthropology, and sociology literatures. Besides realistic network structures and initial conditions, we study three social interaction rules described in these literatures, two of which have been studied in statistical physics as well (random copying or voter rule, and majority rule), and one that has not received as much attention from statistical physicists (following an ``expert'' or a best friend). 

We find that statistical physics framework can be used to develop and compare models of belief dynamics that incorporate a number of plausible social and cognitive mechanisms. When parametrized in line with empirical measures of actual societies (e.g. initial frequency of beliefs and intrinsic preferences, network structure, and relative importance of social information), these models can roughly reproduce real-world patterns of belief dynamics. They enable integration of elements of belief dynamics studied in disparate fields, comparison of the underlying social-cognitive mechanisms across different populations and beliefs, and anticipation of future trends.

The plan of the paper is as follows. In Sect.~\ref{sec:models}, we introduce the basic framework that will be used to model belief dynamics in real-world social networks. In Sect.~\ref{sec:comparison}, we provide examples of models that can be built within the framework and study which features enable them to capture various features of belief dynamics. In Sect.~\ref{sec:modeling} we use the models to predict changes in beliefs and behaviors of individuals and groups observed in two social science studies. We conclude in Sect.~\ref{sec:discussion} by discussing how the results thereby obtained can be used to construct statistical physics models of belief dynamics that, while retaining a relatively simple structure, can nevertheless display both predictive power and shed illumination on the manner in which people use information from their social environments to construct their individual beliefs. 

\section{Models and Methods}
\label{sec:models}

\subsection{Preliminaries}
\label{subsec:prelims}

Given a collection of autonomous entities (usually called ``agents''), a set of states (corresponding to beliefs or behaviors) which the agents can choose among, and a specified dynamics that determines how the agents update their states as time proceeds, one can study how beliefs propagate through a network depending on the details of the model chosen.  In the simplest version, each agent ``lives'' at one of the nodes $i$ of a graph ${\cal G}$, and its state at time $t$ is represented by an Ising spin~$\sigma_i(t)=\pm 1$. That is, each agent has only two possible choices --- agree/disagree, Democrat/Republican, and so on. The ``satisfaction'' of the agent at site $i$ is measured by a Hamiltonian-like function~${\cal H}_i$, consisting of both an external or social, term, and an internal or intrinsic term. Lower values of~${\cal H}_i$ correspond to higher levels of satisfaction. Specifically (for notational simplicity we hereafter suppress the time dependence of $\sigma_i$), 

\begin{equation}
\label{eq:general}
{\cal H}_i=-h_i^{\rm eff}\sigma_i=-h_i^{\rm soc}\sigma_i-h_i\sigma_i\, .
\end{equation}
A model of this type belongs to the general class of random field Ising models (RFIM's), and allows for individual differences among agents. The first term on the RHS of~(\ref{eq:general}) models the overall effect of the social interactions of the agent at $i$, and the second term on the RHS represents the agent's intrinsic predisposition with respect to the belief or behavior being considered. Each agent wants to minimize its value of ${\cal H}_i$,  that is, it wants to align its $\sigma_i$ with its overall effective field $h_i^{\rm eff}$. The overall satisfaction in an entire society consisting of $N$ agents is given by ${\cal H}=\sum_{i=1}^N{\cal H}_i$.

Perhaps the most straightforward (and often-used) representation of the social interaction term is to treat the system as a ferromagnet, i.e., 
\begin{equation}
\label{eq:Hamiltoniani}
{\cal H}_i=-\sigma_i\sum_{j=1}^mJ_{ij}\sigma_j-h_i\sigma_i
\end{equation}
where $m$ is the number of neighbors of $\sigma_i$ (here a neighbor simply means an agent which directly communicates with $i$, i.e., a node connected to $i$ by an edge in ${\cal G}$); the coupling $J_{ij }> 0$ denotes the strength of the interaction between  $\sigma_i$  and $\sigma_j$, and the intrinsic term $h_i$ acts as an individual field acting solely on $\sigma_i$. Both couplings and fields can vary as their indices change, or they can be uniform throughout a given sample. We study different other ways of representing the social interaction term, as described below.

\subsection{Initial conditions for belief distributions}

We first need to determine the specific beliefs whose dynamics we wish to study. These could be, for instance, preferences for different candidates in political elections, attitudes towards different public policies, or beliefs regarding different scientific, health, or economic issues. We can also model adoption of different behaviors, such as dieting, exercise, saving, voting, or paying taxes. 

Once this is determined, a crucial ingredient of any statistical physics model of belief dynamics is specification of the initial condition: what is the initial frequency and spatial distribution of individuals holding different beliefs at the outset? For most of these beliefs and behaviors it is typically possible to at least approximate the initial frequency and spatial distributions of individuals holding different beliefs,e.g. by asking about them in surveys or by inferring them from secondary data sources such as discussions on social media.

 \subsection{The social interaction term}
 \label{subsec:social}  
Studies of humans and other animals have uncovered different rules that individuals use to integrate information from their social environments~\cite{BR85,HL13}. Here we consider three of the most frequently described rules: aligning one's beliefs with a randomly chosen contact, with a majority of contacts, or with a favored person (such as a relative, best friend, or a perceived expert). The first two rules have been extensively studied in the field of statistical physics, while the last one is less well-understood,  as described below. Which rule an individual actually uses depends on the nature of the issue and her individual predispositions, and can be approximated by asking people about their decision-making process~\cite{ELF08}, by fitting different social learning algorithms to their choice data~\cite{MBELR08}, or by conducting laboratory experiments~\cite{MKAN13}. 

\medskip

{\it Voter models.\/}  In these models~\cite{CS73,HL75,CG86,Liggett85,CFL09,FK96,Redner01,SR05,SAR08,Yang09,MGR10,MR11,BCP11,VR12}, an agent updates her belief by choosing one of her neighbors uniformly at random from her connected sites and simply adopts that neighbor's state. This corresponds to a social learning strategy known as unbiased or random copying~\cite{BR85}. In this model, the social field is given simply by
 \begin{equation}
 \label{eq: voterfield}
 h_i^{\rm soc}=\sigma_j
 \end{equation}
where $\sigma_j$ is the randomly chosen neighbor.

\medskip
 
{\it Majority rule models.} In these models, the chosen spin adopts the state of a majority of its neighbors~\cite{Galam02,KR03}. This is equivalent to the ferromagnetic Ising model given by Eq.~(\ref{eq:Hamiltoniani}) when its evolution is governed by zero-temperature Glauber dynamics (i.e., the dynamical rule discussed at the end of Sect.~\ref{subsec:temperature} with $\beta=\infty$).

This rule corresponds to the ubiquitous phenomenon of conformism described in the social learning literature~\cite{BR85,Rendell10} as well as to the majority voting rule that is well-known in political science~\cite{Condorcet}. Because the majority rule doesn't care about the strength of the majority, when $h_i\ne 0$ and/or temperature is greater than zero Eq.~(\ref{eq:Hamiltoniani}) for $h_i^{\rm soc}$ must be replaced by

\begin{equation}
 \label{eq:majority}
 h_i^{\rm soc}={\rm sgn}(\sum_{j=1}^m\sigma_j)
 \end{equation}
where the sum is over the neighbors of the agent at $i$. 
 
 \medskip
 
{\it Expert rule models.\/} Here an agent adopts the belief of one specific neighbor considered to be an expert on the issue (or that otherwise has characteristics that make her/him the favored contact, e.g. being a close friend~\cite{HH13,HL13}). From a modeling perspective, this category of social learning strategies is similar to the voter model, but differs in implementation in that the influential node for each agent is chosen randomly at the start and is thereafter fixed. Similarly to the voter model, the social field here is simply
\begin{equation}
\label{eq: expert}
h_i^{\rm soc}=\sigma_e
\end{equation}
where $\sigma_e$ denotes the state of the chosen expert. 

In all models in this paper, agents applied the social interaction rules on a random sample of three of their neighbors. In this way we can isolate the effect of a specific network structure from the number of neighbors that agents have in different networks. We have chosen three neighbors because this sample size enables easier comparison with our second empirical study described in Sect.~\ref{subsec:Turk} where participants were asked about three of their social contacts. It is also more cognitively realistic as people rarely report consulting with more than two to four close friends about a variety of issues~\cite{Galesic12}.

\subsection{The internal field term}
\label{subsubsec:internal}

The actual updating of the internal state of a chosen agent depends on the interplay between the social interaction term and agent's intrinsic predispositions such as political or religious orientation, lifestyle characteristics, and other general values and behavioral tendencies that can affect the likelihood that a specific belief is accepted. These can be represented as their internal field $h_i$, and their frequency and spatial distributions can be approximated empirically in the same way as specific beliefs: through direct measurement such as surveys and from indirect data such as social media discussions. The relative importance of intrinsic predispositions vs. social influence likely differs for different individuals, societies, and beliefs. We therefore introduce a parameter $0\le \alpha\le 1$, which is the relative weight between the influence of an agent's social network and his individual preferences. A more general formulation than Eq.~(\ref{eq:Hamiltoniani})
for an individual's overall energy or ``satisfaction'' regarding a particular belief is then

\begin{equation}
\label{eq:alpha}
{\cal H}_{i}=-\alpha\sigma_{i}h^{\rm soc}_{i}-(1-\alpha)\sigma_{i}h_{i}
\end{equation}
where $\sigma_{i}$ is individual $i$'s belief state regarding a particular question; $h^{\rm soc}_{i}$ is the social field calculated using social interaction rules such as those described above; and $h_{i}$ is agent~$i$'s intrinsic predispositions to particular beliefs or behaviors.

\subsection{Social network structure}
\label{subsec:network}

Unlike the ordinary statistical mechanics of spin systems, in which networks are usually either finite-dimensional Euclidean lattices or else infinite-range mean-field type structures, models aiming to predict belief dynamics in the real world must consider realistic descriptions of social networks. One frequently used representation are synthetic networks such as ring lattices and fully connected graphs which approximate some real-world networks and (at least for a sufficiently simple Hamiltonian) enable analytic solutions of a system's dynamics. Other more realistic representations include small-world networks or stochastic block models which can reflect properties such as high clustering and short paths, but are typically not analytically solvable. The specific network structure of a particular society can be measured empirically, for instance through sociometric surveys or by automatically recorded data about social contacts~\cite{EP05,Goel10}. Examples of different network structures are provided in Fig.~\ref{fig:lattices} below.

\begin{figure}[!h]
\centering 
\includegraphics[scale=0.60, trim=0 80 20 30, clip]{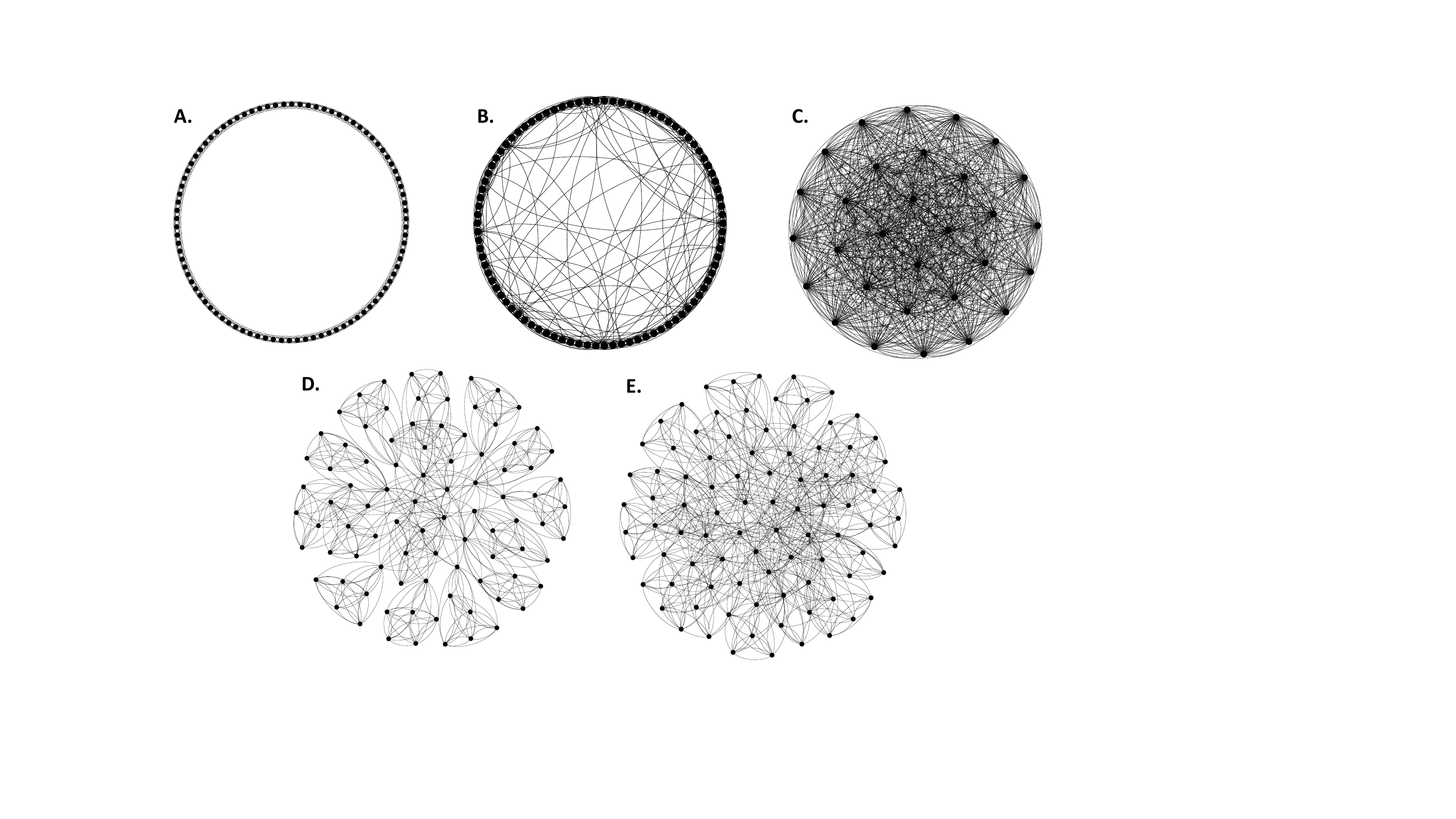} 
\caption{Examples of social network structures investigated in Sect.~\ref{sec:comparison}: A.~ring lattice; B.~small world; C.~fully connected network; D.~weakly connected communities; E.~strongly connected communities.} 
\label{fig:lattices} 
\end{figure}

In the ring lattice (Fig.~1a), each individual is connected to $m$ immediate neighbors. In the small world lattice (Fig.~1b), each node is connected to $m$ immediate neighbors, and also to other nodes with probability $r$ independently for each remaining node. In the fully connected lattice (Fig.~1c), each node is connected to every other node. We compare these stylized networks with more realistic networks generated by stochastic block models. In weakly connected artificial communities (Fig.~1d), individuals are organized in social circles of size $m$; in each circle, one individual is connected to an individual from another circle, leading to at most one connection between any two circles. In strongly connected artificial communities (Fig.~1e), individuals are organized in social circles of size $m$; in each circle, $m/2$ individuals are connected to other circles, leading to multiple connections between any two circles. Detailed properties of each network (including two to be introduced in Sect.~\ref{sec:modeling}) are given in Table~\ref{table:networks} in Appendix~\ref{subsec:networks_details}.

\subsection{Deterministic vs.~stochastic updating}
\label{subsec:temperature}

Finally, we need to specify whether information transfer is perfect or noisy and therefore subject to error. We incorporate this by introducing an adjustable parameter $\beta\ge 0$, which as in statistical physics models behaves as an inverse ``temperature'' of the system. Specifically, $\beta$ represents the reliability of information transfer, or accuracy of an agent's perception of the beliefs of her neighbors. This reliability can be low (small $\beta$) for issues where social information is not easily observable, or when people's beliefs about an issue are prone to random fluctuations (e.g. because they are weakly held or not yet well established). The reliability can be higher when it is clear what others are thinking or doing, and when people's beliefs are strong and well established. 

Once $\beta$ is determined, Eq.~(\ref{eq:alpha}) is used to update the states of the agents. Here we employ asynchronous updating, in which each agent is assigned its own independent Poisson clock, all with rate~$\lambda$. Note that  in the real world it is often reasonable to assume that only a relatively small proportion of individuals change their beliefs at any given time. The updating rate is likely to be different for different beliefs and individuals, but it can be estimated from empirical data. When an agent's clock rings, it evaluates its local energy ${\cal H}_i$. If it can lower its energy ($\Delta{\cal H}_i<0$) by changing its state, it does so with probability one. If changing its state raises its energy ($\Delta{\cal H}_i>0$) , it does so with probability~$[1+\exp(\beta\Delta{\cal H}_i)]^{-1}$. 

Throughout this paper we set $\beta=100$, in which information between agents is transferred with almost perfect accuracy. In future work we will examine cases in which information transfer is noisy or generally less reliable.

\section{Models: Examples}
\label{sec:comparison}

In the previous section we presented a framework that allows development of simple yet cognitively and socially plausible models of belief dynamics that can be directly tested against real-world individual and group data. Before we demonstrate how such tests can be done in~Sec.~\ref{sec:modeling}, we note that this framework can also be used to investigate the effects of introducing different assumptions about social interaction rules, internal field terms, the underlying network structure, the initial conditions, and the relative importance of social and intrinsic factors (parametrized by $\alpha$). We present detailed investigations of these factors in Supplements 1-6~\cite{supp}.

Fig.~\ref{fig:alpha1_and_point5} provides a specific example of how belief dynamics changes as model assumptions vary from less to more realistic. The first two rows show an ideal case where $\alpha=1$, in which social interactions completely govern the belief dynamics, with intrinsic preferences playing no role. The opposite extreme case of $\alpha=0$, in which intrinsic preferences completely govern the belief dynamics and social interactions play no role, can be exactly solved and is discussed in Appendix~\ref{sec:alpha0}. The last three rows show a case where $\alpha=0.5$, meaning that both social interactions and intrinsic preferences play equal roles in belief change. Each panel shows the dynamics occurring if social interaction rule is assumed to be voter, majority, or expert. Within each panel, dynamics for five different networks are shown.

\begin{figure}[!h]
\centering 
\includegraphics[scale=0.75, trim=50 120 20 30, clip] {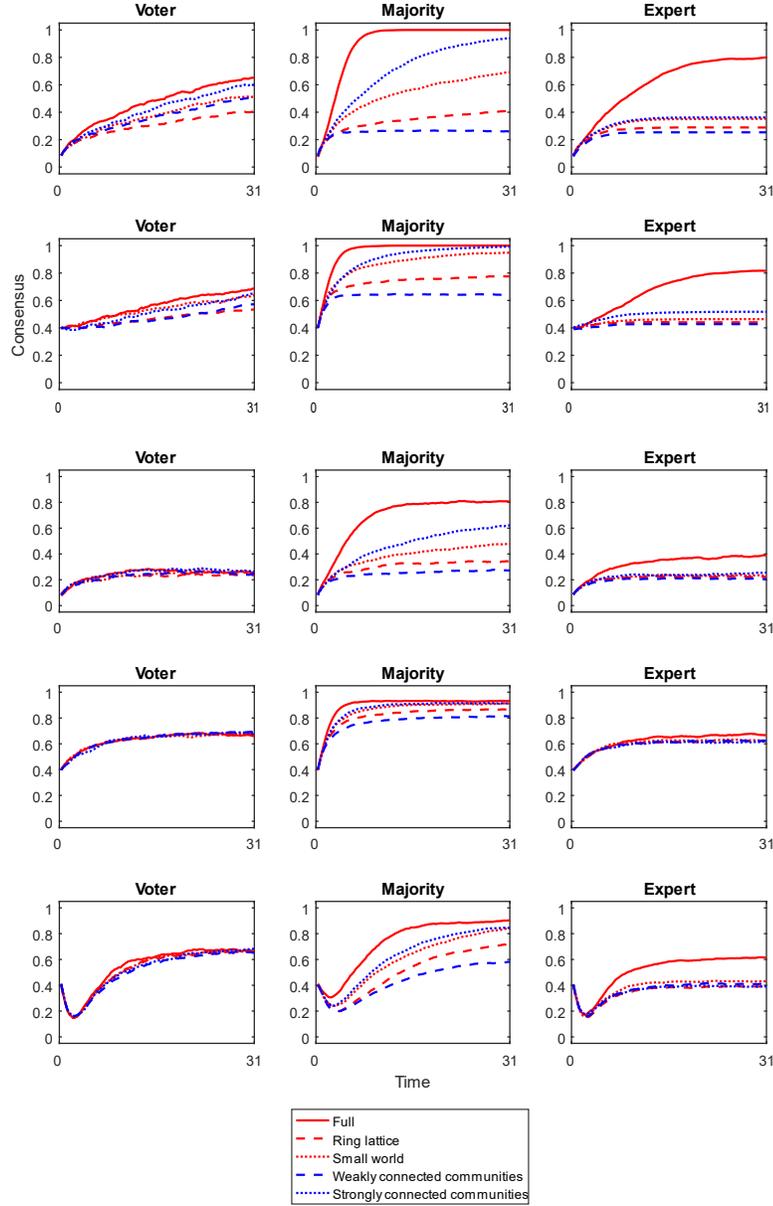} 
\caption{Consensus vs.~time when $\alpha=1$ (rows 1 and 2) and when $\alpha=0.5$ (rows 3-5) for different social interaction rules and different network types, for the two-belief case. Consensus is given by $C(t)=(1/N)|N_1(t)-N_2(t)|$. Row 1 uses a well-mixed initial distribution in which each $\sigma_i=\pm 1$ with equal probability (50-50). Row 2 uses the initial condition where $\sigma_i=+1$ with probability 0.7 and $-1$ with probability 0.3 (70-30). Row 3 uses an initial distribution of 50-50 and $h_i$ drawn from $N(0,1)$. Row 4 uses the initial condition 70-30 and $h_i$ drawn from $N(0.5425,1)$. Row 5 uses the same initial condition but with the $h_i$ drawn from $N(-0.5425,1)$. In all cases $N=90$ and each curve represents an average over 100 runs.} 
\label{fig:alpha1_and_point5} 
\end{figure}

For $\alpha=1$ and an initial distribution of 50-50 (first row of Fig.~\ref{fig:alpha1_and_point5}), we see that network structure is less important for voter rule dynamics than for majority or expert rule. This is hardly surprising, given that the basis for updating an agent's state in the voter rule chooses a single spin at random from the agent's connections, so that the details of network structure are not very important --- all that matters is the number of connections of each agent. 

This is less true for expert rule, where the fully connected lattice (i.e., complete graph) in particular shows a much stronger tendency than other network structures to reach a high degree of consensus. Recall that in expert rule, each agent's reference node for updating remains fixed. As a result, given that in the complete graph every agent is connected to every other, there's a much greater chance (compared to locally connected networks) that ``experts'' will be shared among multiple agents. In contrast, in locally connected networks, only a handful of spins can share the same expert, so the potential for disagreement is greater. Consequently, the more weakly connected a network, the smaller degree of consensus achieved with expert rule. 

Moreover, in locally connected graphs, where geometry plays a role, expert rule can display a peculiar phenomenon absent in the other two dynamical rules. Because experts are chosen at the beginning for each agent and are thereafter fixed, sufficiently large graphs will inevitably display ``closed expert loops'': A's expert is B, B's expert is C, and so on to Z, whose expert is A. A closed loop can be as small as 2: A and B are each other's experts. A closed loop will end in a final state of everyone within the loop agreeing, but which state that will be depends on the initial conditions and the dynamical realization. Hence locally connected graphs with expert rule will display a large degree of metastability with many 1-spin-flip stable states, thereby lowering the overall consensus obtained. For this reason, we expect this behavior to persist for more general cases: expert rule should generally lead to lower values of consensus than the others. This is indeed what is found in our simulations described in Supplement 1~\cite{supp}.

Majority rule exhibits perhaps the most interesting dynamics. Consensus converges toward one for all networks except weak community, which settles at a relatively low rate of consensus (cf. Supplement 2~\cite{supp}). Recall that in the weak community structure, all agents are clustered into small self-reinforcing networks which we call ``social circles'', each with only a single connection to an agent in another circle. Because of the very weak coupling of each social circle to agents outside the circle, with a 50-50 initial condition a given social circle is roughly equally likely to ``freeze'' into either the $+1$ or $-1$ state. That is, there is a large degree of metastability for majority rule that is absent in, say, the voter model. This tendency of majority rule dynamics in which the consensus freezes at relatively low values for certain networks, particularly the weak community structure, persists in many different situations (discussed in Supplement 1~\cite{supp}). For other lattices, one expects convergence to full consensus, which indeed is observed to occur. In fact, this result can be proved for the complete graph (Appendix, Sect.~\ref{subsec:majfull}). 

The above discussion focused on an initial distribution of 50-50. For initial distribution of 70-30 (second row in Fig.~\ref{fig:alpha1_and_point5}), voter rule reaches consensus more slowly than for 50-50 distribution. In contrast, majority rule reaches consensus faster than before (although it is again not reached in weakly-connected communities). Expert rule as before achieves lowest consensus. These behaviors are understood by invoking the same reasoning as used above.

\medskip

We next include the effects of intrinsic preferences: rows~3-5 of Fig.~\ref{fig:alpha1_and_point5} show results for the case $\alpha=0.5$, where social interactions and intrinsic preferences receive equal weight. Row~3 shows the fully ``unbiased case'': i.e., there is a 50-50 distribution of initial condition states, and the internal preferences are symmetrically distributed between $+1$ and $-1$ (modeled by the $h_i$'s drawn from the normal, or Gaussian, distribution $N(0,1)$). The initial conditions and internal fields are both ``well-mixed''; i.e., each initial condition is assigned to an agent independently of all others, and similarly (and separately) for the internal fields. 

An obvious consequence of including sufficiently strong internal fields is that full consensus will no longer be achievable by any social interaction rule or network structure, and this is observed in the third row of Fig.~\ref{fig:alpha1_and_point5}. Another observation is that the influence of the internal fields reduces the dependences on underlying network structure. This is to be expected given that the internal fields each act independently at single sites, so they introduce a network-independent effect, and their strength at $\alpha=0.5$ is statistically equal to that of the social fields. Majority rule remains more sensitive to the underlying network structure than both voter and expert rule, achieving a large degree of consensus ($\approx 0.8$) for the complete graph, and differing degrees of consensus for other lattices: $\approx 0.6$ for strong community, and $\approx 0.4$ for the remaining three (the ring lattice and weak community network don't converge to this final value until longer times than shown in the graph (cf. Supplement 2~\cite{supp}.  

The largest effect is seen for the complete (i.e., fully connected) graph, where every agent interacts with every other. This is not difficult to understand --- such models are mean-field, so that (for sufficiently large $N$) the social field at every site is approximately the same. It should therefore come as no surprise that the degree of consensus obtained here is large compared to local graph structures, and in particular is effective at counteracting the effects of internal fields. A similar phenomenon is observed for expert rule, but because of the sharing of experts among multiple agents in highly connected lattices, as discussed above,
the degree of consensus is significantly suppressed compared to majority rule. For both majority and expert rule, once this unphysical graph is removed, the effects of network structure become less pronounced, and significantly less than for the $\alpha=1$ case. 

For $\alpha=0.5$ and 70-30, initial conditions are aligned with the internal preferences (positive means for the Gaussian distributions of the $h_i$'s; fourth row in Fig.~\ref{fig:alpha1_and_point5}), and all rules and networks achieve consensus faster than before. This is expected given that this case begins close to consensus and the trend toward consensus is reinforced by the alignment of initial beliefs and internal preference fields. The more interesting case is that where the initial condition is 70-30, and the internal preference is 30-70 ($h_i$'s drawn from $N(-0.5425,1)$), that is, opposite to the initial condition. Results are shown in the fifth row of Fig.~\ref{fig:alpha1_and_point5}. Because most internal fields are aligned, they force realignment of beliefs to be more in accordance with each agent's internal preference, slowing down the consensus compared to the aligned scenario above. The behavior in rows 4 and 5 (in particular the initial dip in curves shown in the fifth row) can be understood by referring to the exact solution of the $\alpha=0$ case, presented in Appendix~\ref{sec:alpha0}.

The framework described in Sect.~\ref{sec:models} allows for construction of many further models that can mimic situations with varying degrees of social influence compared to intrinsic preferences ($0<\alpha < 1$). We show these results in Supplement 2~\cite{supp}. The framework also allows for investigation of models that assume spatially clustered internal field distributions (Supplement 3), larger networks (Supplement 4), four (Supplement 5) or five (Supplement 6~\cite{supp}) beliefs instead of just two, and so on. 

Taken together, results presented in this section suggests that accurate modeling of real-world patterns requires attending to several properties of real-world societies, in particular spatial and frequency distributions of internal predispositions, initial conditions, social interaction rules, and network structures. However, to avoid overfitting one should ideally determine the values of as many of these parameters as possible at the outset by empirical measurement in a particular society, or from theoretical understanding of human behavior, rather than determine them separately for each study or application. We demonstrate this approach in the next section. 

\section{Modeling belief dynamics in real-world societies}
\label{sec:modeling}

We turn now to two fairly detailed studies on belief (and behavior) dynamics in two different kinds of social networks, and investigate whether the models described above can be useful in understanding the mechanisms underlying belief dynamics in these real-world settings. 

\subsection{Modeling belief change in the MIT Social Evolution Project}
\label{subsec:MIT}

\subsubsection{Procedures}
\label{subsubsec:MITprocedures}

In 2008 and 2009, a study was conducted on 80 students and student tutors who lived in an undergraduate dormitory at MIT; details of the study and its results can be found in~\cite{MCMFP12} and at 
http://realitycommons.media.mit.edu/socialevolution.html. The data consisted of successive survey waves which measured the participants'  beliefs and behaviors, as well as their network characteristics.

\medskip

{\it Beliefs.\/} Initial distributions of individual beliefs and behaviors were determined from participants' answers in the first survey wave (September 2008) to questions about politics and health. These included:  

\begin{enumerate}

\item {\it Political beliefs.\/} Survey questions in this area focused on likelihood to vote, preferred candidate,  overall interest in politics, preferred party, and political orientation as of September 2008.  Responses were collected in two survey waves before the 2008~presidential election (September and October 2008) and in one survey wave afterward (November 2008).

\medskip

\item {\it Health-related behaviors.\/} Participants reported on the number of salads they ate per week, portions of fruits and vegetables per week, number of days per week when they did at least 20~minutes of aerobic exercise, number of sport activities per week, and overall healthiness of diet. These data were collected in survey waves conducted in September, October, and December of 2008, and March, April, and June of 2009. Here we restrict ourselves to the questions about consumption of fruits and vegetables and sport activities (the other two questions showed similar patterns).

\end{enumerate}
\bigskip

Table~\ref{table:MITquestions} contains details of survey questions, the possible responses to each question, and the way responses were coded for modeling purposes. The number of possible beliefs and behaviors differed depending on the issue. Note that for political questions the number of possible responses was larger before the presidential election (September and October~2008) than after the election (November~2008). To account for this and enable predictions over time, we recoded all answers in September and October as follows: likelihood to vote ranged from -1 (including answers 'definitely will vote' and 'most likely will vote') to 1 (including answers 'don't know', 'most likely will not vote', 'definitely will not vote'). Preferred candidate response values ranged from -1 (including 'definitely Obama' or 'probably' Obama), to 0 ('other candidate' or 'undecided' or in November, 'did not vote'), to 1 (including 'definitely McCain' and 'probably McCain').

\begin{table}[h]
\centering 
\includegraphics[scale=0.60]{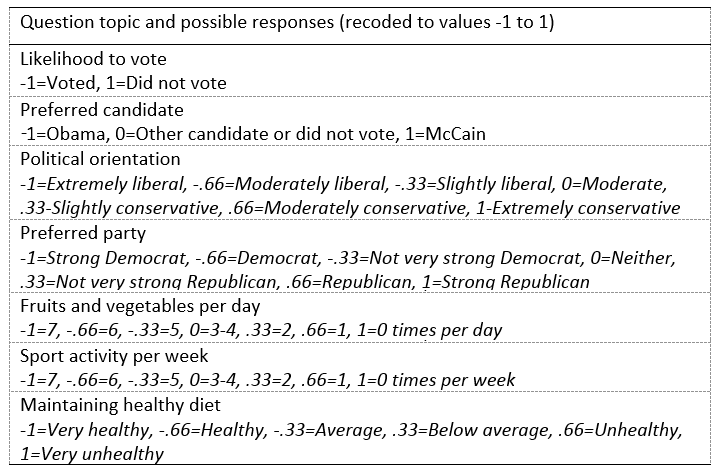}
\caption{Questions in the MIT study.}
\label{table:MITquestions} 
\end{table}

\medskip

{\it Networks.\/} We compared three artificial network structures that are often used in the statistical physics literature to mimic the real-world networks (fully connected, ring lattice, and small-world networks) with two empirically derived networks described below. This comparison enabled us to evaluate how well these artificial networks represent the real-world society studied here.

\begin{enumerate}

\item  {\it Objective networks\/} were determined through proximities of individuals as captured by Bluetooth signals indicating that the cell phones of two individuals were within 10 meters of each other. These data were additionally weighted by the likelihood that the two individuals were on the same floor, as captured by scanning nearby WiFi access points. For these we use data for the months October-December 2008, and March-June 2009.

\medskip

\item {\it Subjective networks\/} were determined by collecting sociometric data from individuals about their contacts and the nature of their relationship with each contact (close friend, political discussant, someone they socialize with twice per week, and/or social media contact). Edges are weighted by the strongest relationship between two people, with edge values ranging from 1 for close friends (or a person with whom one discusses politics, for political questions), .8 for people who socialize at least 2 times per week, to .5 for people who share their tagged Facebook photos, blog, live journal, or Twitter activities. For these we use data from survey waves conducted from October to December 2008, and from March to May 2009.

\end{enumerate}

\begin{figure}[!h]
\centering 
\includegraphics[scale=0.5, trim=0 300 20 150, clip]{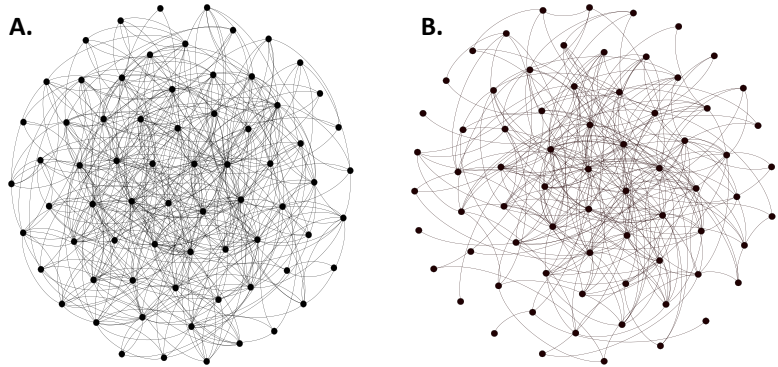} 
\caption{Empirically derived network structures investigated in the MIT study: A. objectively measured network; B. subjectively measured network.  Characteristics of these and other networks examined in this paper appear in Table~\ref{table:networks}.}
\label{fig:obj} 
\end{figure}

The objective and subjective networks are shown in Fig.~\ref{fig:obj}. The edges change somewhat at each time point, reflecting the development of social relationships in this particular society over time. Detailed characteristics of these networks can be found in Table~\ref{table:networks} in Appendix~\ref{subsec:networks_details}. 

\medskip

{\it Internal fields. \/} We assumed that the relevant intrinsic predispositions can be inferred from participants' answers to questions that reflected their broader views about politics (for likelihood to vote: interest in politics; for preferred candidate: average of preferred party and political orientation) and health (for both fruits and sports: trying to maintain a healthy diet). 

\medskip

{\it Social interaction term. \/} We investigate how well can the empirical data be explained by assuming that participants used each of the three social learning rules discussed above: voter, majority, or expert/best friend rule. Because we do not have empirical measures of the rules that participants actually used, we compare predictions based on each of the three rules to see which one predicts the empirically observed belief dynamics best. Both voter and majority rule are straightforwardly applied: voter rule entails taking a social signal from a randomly chosen member of one's network contacts, while majority rule takes as a social signal the belief or behavior of the majority (or plurality if there are several options) of one's contacts. For expert/best friend rule, we used for each individual a contact whose belief one might consider to be particularly relevant for the decision under question. For political beliefs, we considered an expert to be a person that is either a close friend or a person that the individual discusses politics with. For health-related behaviors, an expert is considered tobe one's close friend, who is presumably both similar to and attuned to the individual's own dietary and physical activity needs and preferences. 

\medskip

{\it Belief updating. \/} All three social rules assume that people first select a random sample from their contacts, weighted by frequency of contact (for objective networks) or closeness of contact (for subjective networks), and then apply the particular rule on that sample. As in Sect.~\ref{sec:comparison}, we then use the Hamiltonian~(\ref{eq:alpha}) to model the time evolution of belief distribution throughout the networks. The only difference is that here we go beyond the basic Ising model that accepts only two beliefs and apply a version of the Potts model that allows for multiple choices. 

Modeling proceeded as in Sect.~\ref{sec:comparison}: at each time step (i.e., survey wave), an individual is chosen with probability given by a Poisson distribution with rate~$\lambda$ of updating. If chosen, the individual determines his or her current social-cognitive dissonance~${\cal H}_i$ according to Eq.~(\ref{eq:alpha}), and (if possible) changes his or her belief/behavior to that which would best reduce the dissonance.

{\it Parameter estimation.\/} The model has two free parameters: the rate of updating~$\lambda$, and the weight~$\alpha$ assigned to social information. We determined the value of these two parameters for each combination of network structure and updating rule, using grid search in steps of 0.1 from 0 to 1. We selected the combination of parameter values that produced the best predictions (in terms of average percentage of correct individual predictions) of beliefs at the second time point (October) from beliefs at the first time point (September). To avoid overfitting given the small number of data points for each participant, we fixed the values of~$\lambda$ and~$\alpha$ at their average value across all networks and used these values to predict beliefs and behaviors for all future time points and for all participants. For each combination of assumptions, this procedure is repeated 50 times and the results are averaged across replications. Note that in each replication we assume a different set of connections in two incompletely connected artificial networks, ring lattice and small world. As before, inverse temperature is fixed at $\beta=100$. 

\subsubsection{Results}
\label{subsubsec:MITresults}

We begin by examining parameter values determined to produce the best predictions of October~2008 beliefs and behaviors, based on participants' answers in September 2008. These parameter values are presented in Table~\ref{table:parameters}. The updating rates are quite low, suggesting that only a minority of participants updated their beliefs in each survey wave. Also unsurprisingly, the weight of social information is higher than the weight of internal preferences for all questions except preferred presidential candidate, where intrinsic political orientation seems to play a stronger role than social interactions (at least in this relatively homogeneous, predominantly liberal community). It is particularly interesting that the highest weight given to social interaction corresponds to an individual's likelihood to vote, where one might expect social pressure could play an important role in a close-knit community. The next highest weight for social interaction corresponds to sports played per week; given that many sports are group activities, this seems plausible and speaks to the utility of the models.

\begin{table}[h]
\centering 
\includegraphics[scale=0.70]{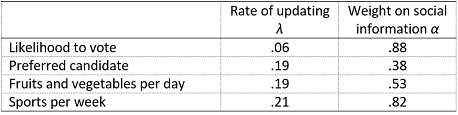}
\caption{Values of parameters $\lambda$ and $\alpha$, fitted by grid search for the combination that produced the best prediction of answers at the second time point (October 2008) based on answers at the first time point (September 2008), in the MIT study. These average parameter values were then used to predict participants' answers at the third time point and beyond.}
\label{table:parameters} 
\end{table}

\begin{figure}[!h]
\centering 
\includegraphics[scale=0.6, trim=0 100 0 0, clip]{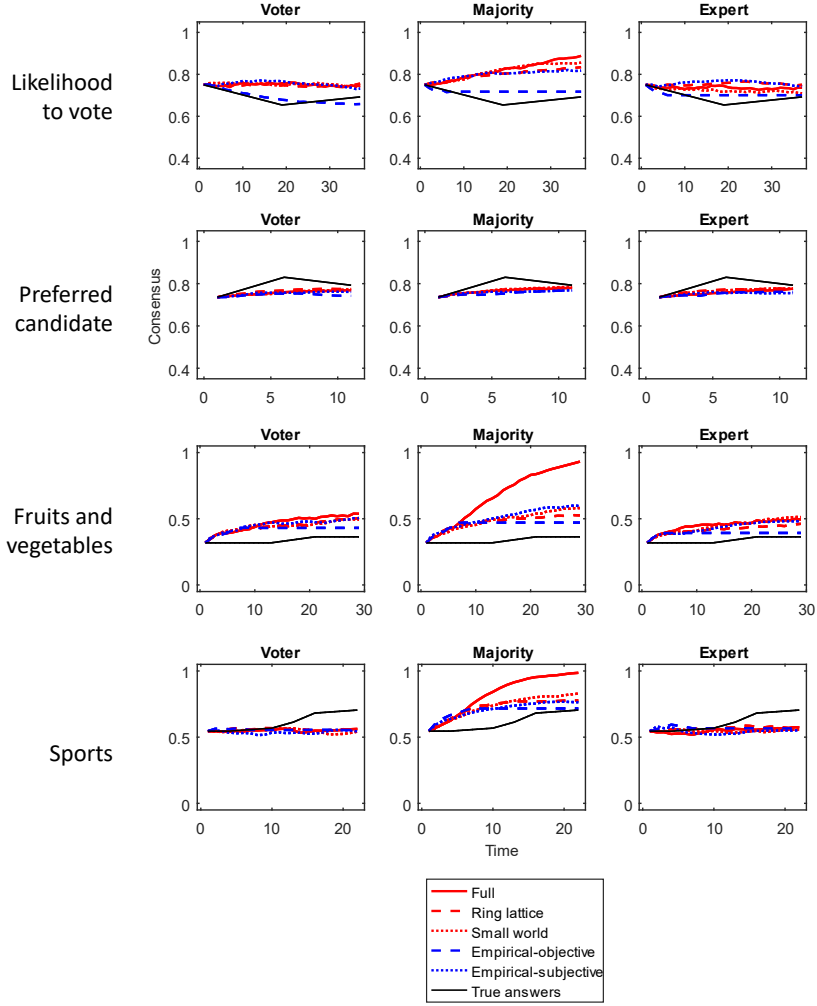} 
\caption{Empirically observed and modeled level of consensus, for questions in the MIT study, using different assumptions about social learning rules (columns) and underlying network structures (lines, see legend). The results are averaged across 50 replications.}
\label{fig:MIT_consensus} 
\end{figure}

We now turn to model predictions for specific questions. Fig.~\ref{fig:MIT_consensus} shows predictions for changes in the level of consensus over time for the political and health-related questions. Most predictions are reasonably close to the true level of consensus, given that the models are very simple and use the same values of~$\lambda$ and~$\alpha$ for all networks and rules. However, looking at the level of consensus alone is not sufficient: even if it is predicted well, it could be that the models predict convergence on the candidate who receives a minority of the vote. Group-level patterns of consensus formation will look the same, whether the group converges on one particular belief or on the completely opposite belief. The exact belief that the society is converging on makes little difference in abstract models, but is important if the model is going to be useful in the real world. 

We therefore complement Fig.~\ref{fig:MIT_consensus} with a figure showing average proportion of different responses (Appendix~\ref{subsec:empiricalfigs}, Fig.~\ref{fig:MIT_answers}), and a table showing the average accuracy of predictions on the individual level (Table~\ref{table:Table_RMSD_MIT}). The accuracy is measured as the square root of the average of the squared difference (RMSD) between the actual and predicted answers of each individual, averaged across all time periods but the first one, and across 50~replications. The accuracy is highest when the RMSD is 0, and lowest when it equals 2.

\begin{table}[h]
\centering 
\includegraphics[scale=0.80]{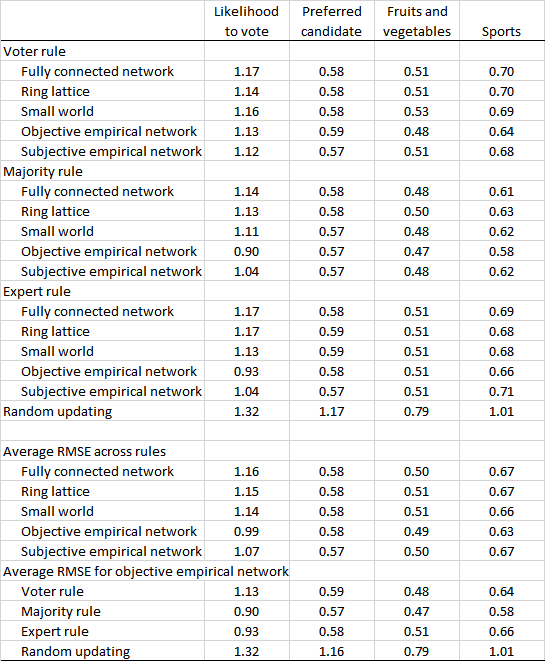}
\caption{Individual-level prediction accuracy of different models, for questions in the MIT study, expressed as root mean squared deviations (RMSDs) between predicted and true answers. Lower values are better. The top part shows performance of each rule in different networks. As a baseline, we include results that would be obtained by random choice of answers ("random updating"). The bottom part shows average performance of each network and performance of different rules for the best performing network. Note that RMSDs for different questions cannot be directly compared because questions had different number of response options (see Table~\ref{table:MITquestions}).}
\label{table:Table_RMSD_MIT} 
\end{table}

Models using different assumptions produce similar results, but we can nevertheless observe some qualitative regularities occurring across questions, networks, and updating rules. The results suggest that empirical networks do better than the theoretical ones, with objectively measured empirical network performing somewhat better than the subjectively measured ones. The fact that the objectively measured network outperformed the subjectively measured one suggests that people's own reports about their network connections do not include all relevant interactions, which are recorded by automated measurements. 

Given that the results point to the objective network as the most realistic, it makes sense to compare the three social interaction rules by focusing on the objective network only; this removes the noise generated by modeling these rules on less realistic networks. The bottom part of Table~\ref{table:Table_RMSD_MIT} presents the error of predictions obtained assuming different social interaction rules and the objectively measured network. 

All three social interaction rules outperform the baseline assumption of random updating. For most questions majority rule performs somewhat better than expert rule, and both do better than voter rule. This is encouraging, given that voter rule (without individual fields) was formulated as a tractable analytical model but perhaps should not be expected to function well in a realistic setting. Presumably, few people take on the belief of a randomly selected contact. The fact that majority and expert rule indeed do a better job at prediction is a validation that the relatively simple statistical physics approach studied in this paper does capture some aspects of belief dynamics in the real world. It further provides evidence for the plausible conjecture that in formulating beliefs, people rely on their intrinsic preferences, belief of a majority of their neighbors, and sometimes also the belief of an ``expert''.

\subsection{Modeling belief change in a longitudinal survey study of the US public}
\label{subsec:Turk}

\subsubsection{Procedures}
\label{subsubsec:turkprocedures}

For the second comparison of models to data, we conducted a longitudinal study in four survey waves, each two weeks apart, with participants recruited from the crowdsourcing service Mechanical Turk. Two~hundred participants participated in the first wave, and 94 completed all four waves of study. The study was conducted in a relatively turbulent political period in the United States, with the first two waves conducted just before the first 2016~presidential primary (Iowa, February 1), and the other two afterward. 

{\it Beliefs.\/} In each wave, participants were asked about several topics. Distribution of initial beliefs was determined based on participants' responses in the first survey wave. For simplicity, we present here only a subset of the questions and the results pertaining to them; these illustrate well the different observed trends. The questions reported here were taken from existing national surveys and covered the topics of guns, terrorism, and vaccinations; their exact wording, possible responses, and the coding of responses appear in Table~\ref{table:Turkquestions}. Information was also collected on additional characteristics of participants' social contacts, including their perceived expertise about the issues, their preferred social learning rules, other sources of information they use, rules for network updating, and several psychological traits. A full description with the complete set of questions and trends will appear elsewhere~\cite{GSinprep}.

\begin{table}[h]
\centering 
\includegraphics[scale=0.55]{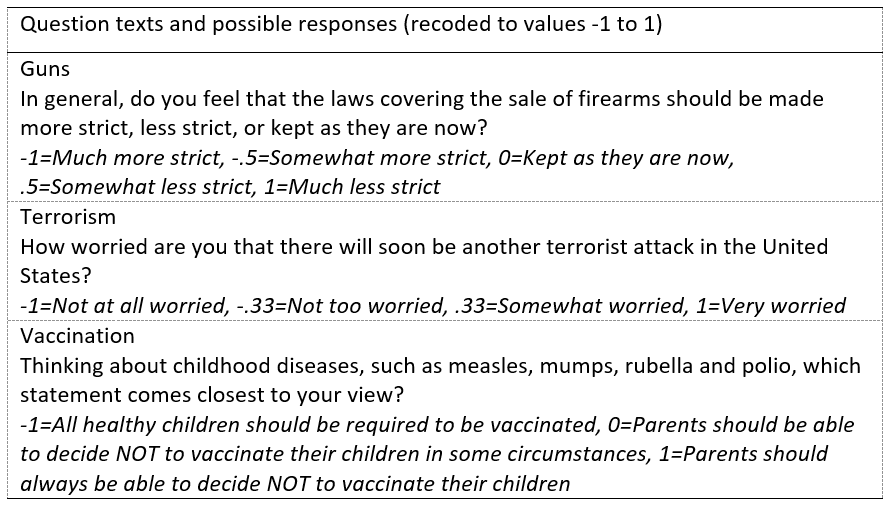}
\caption{Questions in the Mechanical Turk survey.}
\label{table:Turkquestions} 
\end{table}

\begin{figure}[!h]
\centering 
\includegraphics[scale=0.5]{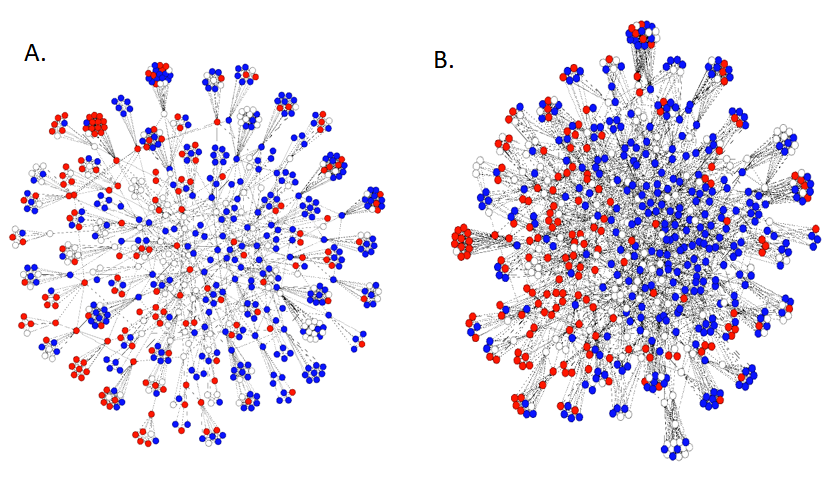}
\caption{Empirically derived network structures investigated in Sect.~\ref{subsec:Turk}: a.~weakly connected communities; b.~strongly connected communities. Connections between the communities in the two networks are based on empirical patterns of connections between participants with different political orientations (blue=left; red=right; white=center). See text for more details.}
\label{fig:Turk_networks} 
\end{figure}

{\it Networks.\/} Unlike in the MIT study, members of the participants' social network did not participate in the survey themselves. We therefore needed to rely on the participants' perceptions of beliefs held by their social contacts. Participants were asked to report on the perceived beliefs, on each of the issues, of up to three of their social contacts with whom they discussed these issues in the two weeks prior to the survey. These social contacts could differ from wave to wave. On average, each participant reported about 5.8~contacts across the four study waves, for a total of 548~different contacts. To construct empirical networks, we assumed that our 94 participants were embedded in the social circles composed of their reported contacts, and also connected with one (for the weakly connected communities) or two (for the strongly connected communities) of the other 93 participants. The probability of a link between any two participants was determined from the data as the probability that two people in the same social circle share the same political orientation. The resulting network structures reflect the assumption that people with similar political orientation tend to be exposed to similar information sources and therefore are affected by new events in similar ways. We compared these empirical network structures, shown in Fig.~\ref{fig:Turk_networks}, with the same three artificial networks as in the MIT study (fully connected, ring lattice, and small-world networks).

{\it Internal fields.\/} These were operationalized as participants' political orientation, reported on a scale ranging from extreme left to extreme right.

{\it Social interaction term.\/} For voter rule, we used the perceived belief of a random contact; for majority rule, the perceived belief of a plurality of a participant's contacts; and for expert rule, the perceived belief of the contact determined by the participant to have the greatest expertise.  

{\it Belief updating.\/} As in the MIT study, we use a Potts-like model that starts from the initial belief distribution measured in the first wave, and develops predicted belief trajectories at future time points using predicted beliefs in each subsequent wave. For each combination of factors (five network structures and three social learning rules), we ran 30~replications lasting 160~time steps, which was enough to achieve stable results. 

{\it Parameter estimation.\/} To to test how well parameters estimated for one group translate to a wholly different group of participants (out-of sample prediction), we used a different procedure from that used in the MIT study to determine the parameters~$\lambda$ (rate of belief updating) and $\alpha$ (relative weights between social information and intrinsic preferences). Rather than using first-wave data to determine these parameters separately for each question, we instead used the average values from the previous study (presented in Table~\ref{table:parameters}); consequently, $\lambda$ was set to~0.17 and $\alpha$ to~0.67 for all participants and all questions. 

\subsubsection{Results}
\label{subsubsec:Turkresults}

\begin{figure}[!h]
\centering 
\includegraphics[scale=0.6, trim=30 200 0 30, clip]{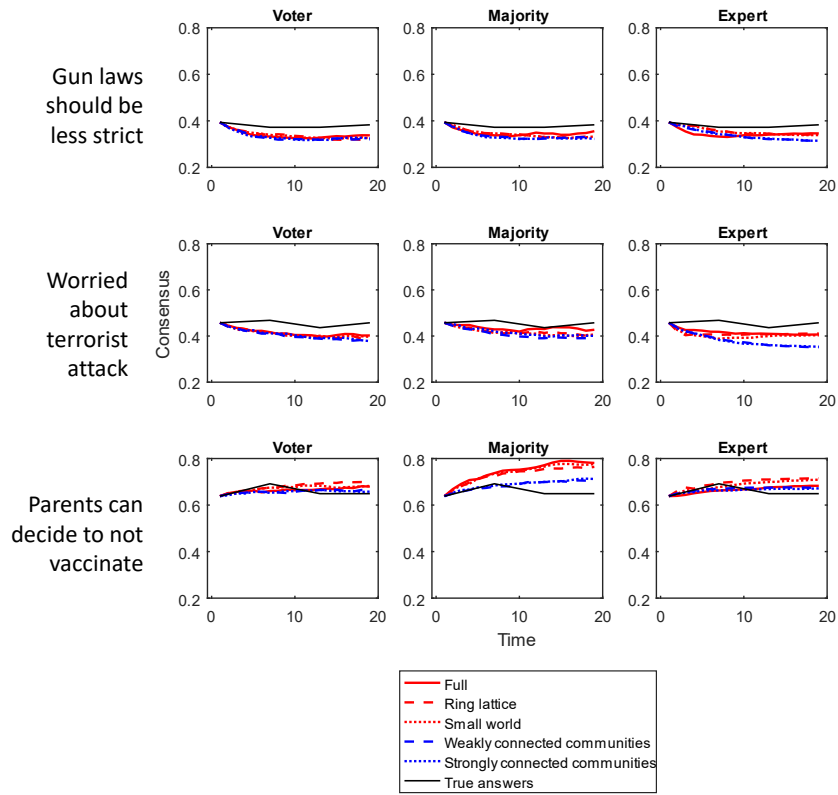}
\caption{Dynamics of consensus in the Mechanical Turk study: model predictions assuming different underlying network structures and social learning rules, compared with the actual answers. The results are averaged across 50 replications.}
\label{fig:Turk_consensus} 
\end{figure}

Fig.~\ref{fig:Turk_consensus} shows empirical trends of belief change related to the three issues across the four study waves, with predictions obtained using Eq.~\ref{eq:alpha} in three different network structures. Model predictions assuming different network structures and social learning rules correspond reasonably well to empirical trends, considering their simplicity and the fact that the parameters were not fitted to the present data but generalized from the previous study. As before, we complement Fig.~\ref{fig:Turk_consensus} with a figure showing average proportion of different responses (Appendix, Fig.~\ref{fig:Turk_answers} in Sect.~\ref{subsec:empiricalfigs}), and a table showing the average accuracy of predictions on the individual level (Table~\ref{table:Table_RMSD_Turk}). Table~\ref{table:Table_RMSD_Turk} shows average deviations of predicted from true individual answers for different network structures.

The two empirical network structures (weak and strong communities) produce much better predictions than the three synthetic networks (full, ring, and small world). This suggests that there is a value in the approach of, whenever possible, using empirically derived network structures rather than trying to approximate them using artificial networks. Restricting our attention to the relatively best performing weakly-connected community networks, we again observe that on average the majority and expert rules do better than voter rule, and all do better than random prediction. Differences in rule performance across questions make sense: for instance, expert rule performed particularly well relative to the majority rule for the question about vaccination, suggesting that at least in this study most participants form their beliefs about this issue based on their doctors' advice.

\begin{table}[h]
\centering 
\includegraphics[scale=0.80]{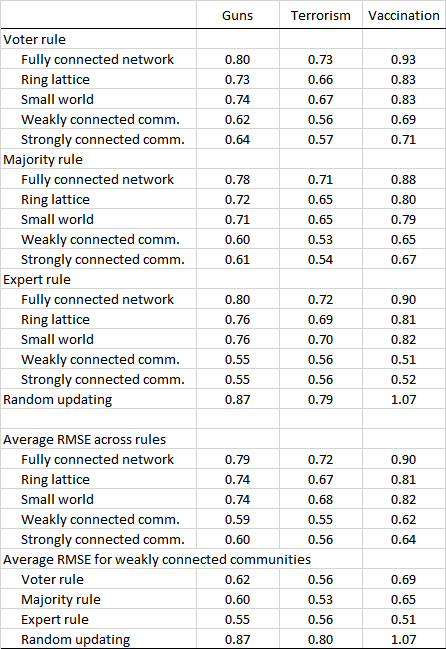}
\caption{Individual-level prediction accuracy of different models, for questions in the Mechanical Turk study. Accuracy is expressed as root mean squared deviations (RMSDs) of predicted and true answers. Lower values are better. The top part shows performance of each rule in different networks. As a baseline, we include results that would be obtained by random choice of answers ("random updating"). The bottom part shows average performance of each network and performance of different rules for the best performing network. Note that RMSDs for different questions cannot be directly compared because questions had different number of response options (see Table~\ref{table:Turkquestions}).}
\label{table:Table_RMSD_Turk} 
\end{table}

\section{Discussion}
\label{sec:discussion}

There is now a large and continually growing literature dealing with the application of statistical physics models to belief dynamics in social networks. Beginning with the homogeneous 2-state voter model on simple Euclidean lattices, these models have increased in variety, complexity and sophistication. It has consequently become difficult to evaluate similarities and differences in behavior of these many approaches, as well as their relative merits in explaining and predicting belief propagation in actual human communities. While comparisons to empirical data, whether gleaned from elections or from controlled social studies, have appeared in the literature, they focus only on group- rather than individual-level patterns and remain relatively rare compared to the proliferation of new models and approaches.

We present a framework that can be used to implement a variety of plausible assumptions about social-cognitive mechanisms underlying belief dynamics. The framework has the structure of well-studied discrete-spin models, which helps constrain the number of variables in the model, as well as anticipate basic dynamics resulting from different modeling assumptions.

In these models, agents can adopt one of a small number of available states while interacting in different network structures. These models constitute the largest subset of the literature of physics-based social science models. Most models studied so far are homogeneous, that is, every agent is assigned the same Hamiltonian-like function (a subset of these are mean-field models). In contrast, here we we have added a random-field term to account for the empirical observation that people differ in their propensity to accept different beliefs. 


As discussed in Sect.~\ref{sec:models}, even within this limited class of models there are many possible variations. There are numerous possibilities for the social interaction term, internal field distribution, initial conditions, social network structure, rate of updating, relative weights of social information vs.~intrinsic preferences, and so on. There is also the question of deterministic vs.~stochastic updating of beliefs. In Sect.~\ref{sec:comparison} we presented several examples of how introducing more realistic assumptions about these factors can produce very different belief dynamics. This variety of options reinforces the need to parameterize such models using empirical data about a particular society, rather than trying to explore the vast space of plausible models.

Our modeling results suggest that among the three social interaction possibilities, the voter model is the least sensitive to network structure and initial conditions. It is also typically the slowest to arrive at the final consensus value; although it can ultimately achieve high consensus values. For majority and expert rules, final consensus values were much more sensitive to both network structure and initial condition distribution. Our comparison to real-world data confirms that this sensitivity to network structure is indeed an important factor in determining the degree of consensus.

Perhaps one of the most important conclusions arising from the studies in Sect.~\ref{sec:comparison} is that the most realistic network structures, social learning rules, and spatial distribution of beliefs (as determined in empirical studies such as those described in Sect.~\ref{sec:modeling}) are those least likely to achieve high levels of consensus. This  corresponds to real-world behavior and contrasts to those statistical physics models which achieve high levels of consensus. A summary of other conclusions arising from comparisons of different models can be found in Supplement 1~\cite{supp}.

To constrain the large number of possible combinations of seemingly plausible parameters, we go beyond modeling possible societies and use empirical data from two survey studies to parametrize and evaluate our models. As described in Sect.~\ref{sec:modeling}, the first was a study of 80~students in an MIT dormitory during the presidential election of 2008, and the second an online  study with 94~participants conducted by us during the presidential election of 2016.  The MIT~study looked at  political beliefs and health-related behaviors, while the longitudinal study focused on politics and science-related issues.

Comparing the data obtained in each of these studies with predictions made by the various discrete-spin models, we found that models using objectively determined network structures did a better job of capturing the data than those using artificial networks (Euclidean lattices, complete graphs, small world, ring lattices, and others). Moreover, majority and expert rule generally did much better than voter rule, with the precise nature of the question determining whether majority or expert rule had more predictive value. 

Our empirical studies led to two main conclusions. First, discrete spin models do have real predictive value despite their simplicity, and even in the worst cases do better than control conditions using random prediction. Second, when one looks at globally averaged behaviors one finds similar aggregate trends that are not very sensitive to different social learning rules, underlying network structures, and spatial distribution of beliefs. In contrast, individual-level trends are often very sensitive to these factors. This underlines the importance of understanding belief formation at the individual level, and supports the importance of modeling individual differences in social interaction terms, internal fields, rate of updating, and weights of social information vs.~intrinsic preferences. In future studies, we will therefore investigate the dynamics exhibited by populations of agents that use a mixture of different social interaction rules, to approximate the heterogeneity observed in the real world. We will also investigate whether the inclusion of noise (positive-temperature models), in which an individual's perceptions of others' beliefs may contain errors, helps to capture real-world belief dynamics.

More generally, our results suggest that simple computational models constructed using a statistical physics approach are able to produce patterns that resemble real-world dynamics, and as such can be used as a convenient framework for building tractable yet cognitively and socially plausible computational models of belief dynamics that can illuminate important aspects of the underlying mechanisms. Although there are many excellent agent-based models of social phenomena, they tend to differ for each new phenomenon and authors have a lot of discretion in deciding what parameters to include. Statistical physics models can provide one way to determine which parameters are necessary and sufficient to reproduce the empirically observed dynamics (for another approach, see~\cite{Epstein14}). They can also help to organize a number of currently disparate social and cognitive concepts that are involved in belief dynamics, including initial frequency and spatial distribution of beliefs and intrinsic predispositions, structural properties of social networks, perceptual noise, and cognitive rules for social interactions and belief updating. A number of issues remain to be resolved, including how these different concepts can best be operationalized and measured, and how they relate to existing models of belief updating and dynamics in cognitive science, social psychology, and sociology. These issues will be the topics of future work.

\begin{acknowledgements} 

MG was supported in part by NSF MMS-1560592 and by NSF SMA-1633747. DLS was supported in part by NSF DMS-1207678. Any beliefs, findings, and conclusions or recommendations expressed in this material are those of the authors and do not necessarily reflect the views of the National Science Foundation. DLS thanks the Guggenheim Foundation for a fellowship that partially supported the early stages of this research, and the Santa Fe Institute and Aspen Center for Physics (under NSF Grant 1066293), where some of this work was carried out.  We thank Luis Bettencourt, Henrik Olsson, and Sid Redner for reading through a preliminary version of the paper and offering helpful comments.

\end{acknowledgements}

\bibliographystyle{unsrt}

\bibliography{refs}

\clearpage

\appendix
\label{sec:app}

\section{Properties of networks}
\label{subsec:networks_details}

\begin{table}[h]
\centering 
\includegraphics[scale=.40]{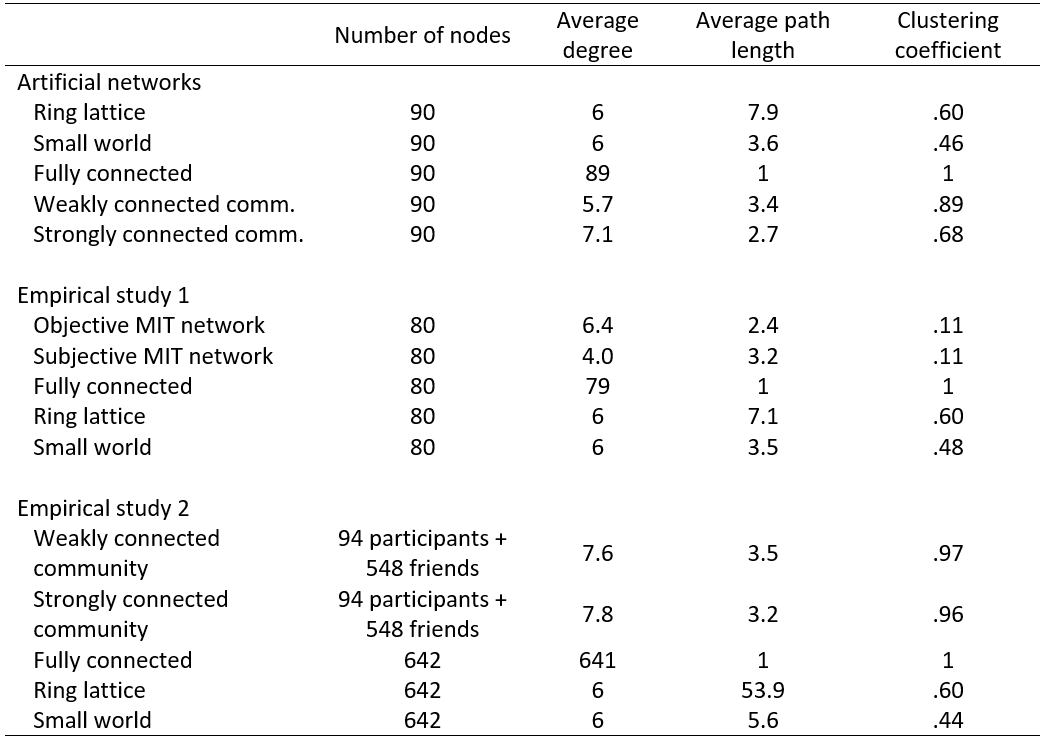}
\caption{Properties of networks studied in this paper.}
\label{table:networks} 
\end{table}

\section{The case of internal preferences only: $\alpha=0$}
\label{sec:alpha0}

In this Appendix we discuss the case $\alpha=0$, where social interactions have been turned off. This case is exactly soluble analytically, so simulations are not needed (except as a consistency check). Because the social interaction term is absent, the graph on which the agents sit is irrelevant; all that matters are the initial conditions.

When $\alpha=0$, the Hamiltonian is simply ${\cal H}=-\sum_i h_i\sigma_i$. The zero-temperature process for the case of two possible beliefs can then be modeled as a 4-state Markov chain. Let $N_{11}(t)$ denote the number of agents in state 1 at time $t$ and with $h_i=+1$ (i.e., they prefer option 1); $N_{12}(t)$ is the number of agents in state 1 at time t but with $h_i=-1$ (they prefer option 2); and similarly for $N_{21}(t)$ and $N_{22}(t)$. The $h_i$'s, which are fixed in time, are uniformly distributed Bernoulli random variables (however, the analysis and results below hold for any well-defined distribution of the $h_i$'s that is symmetric about zero), and the initial states $N_{11}(0)$, $N_{12}(0)$, etc., are also uniformly distributed according to some initial distribution. Of course, $N_{11}(t) + N_{12}(t) + N_{21}(t) + N_{22}(t) = N$, independent of time.

If an agent is in the state $\{11\}$ or $\{22\}$ at any time, it thereafter remains in that state. If it's in the state $\{21\}$, then when its Poisson clock rings it changes its state from $-1$ to $+1$ with probability $p$; similarly, if it's in the state $\{12\}$, it changes its state from $+1$ to $-1$ with probability $q$. 

The dynamical equations for the chain are therefore (in a discrete time formulation)
\begin{equation}
\label{eq:matrix}
\left[\begin{array}{c}
N_{11}(t)\\
N_{12}(t)\\
N_{21}(t)\\
N_{22}(t)\end{array}\right]=\left[\begin{array}{cccc}
1&0&p/N&0\\
0&1-q/N&0&0\\
0&0&1-p/N&0\\
0&q/N&0&1\end{array}\right]\left[\begin{array}{c}
N_{11}(t-1)\\
N_{12}(t-1)\\
N_{21}(t-1)\\
N_{22}(t-1)\end{array}\right]\, .
\end{equation}
Letting ${\bf N}(t)$ denote the 4-component state vector at time $t$ and
${\bf A}$ the $4\times 4$ Markov transition matrix, Eq.~(\ref{eq:matrix}) can be written compactly as ${\bf N}(t)={\bf A}{\bf N}(t-1)$; it follows that ${\bf N}(t)={\bf A}^t{\bf N}(0)$.  This set of equations is easily solved by diagonalizing ${\bf A}$; rescaling $t$ by $Nt$ and letting $N\to\infty$, the solution is
\begin{eqnarray}
\label{eq:solution}
N_{11}(t) = N_{11}(0) + N_{21}(0)\left(1 - e^{-pt}\right)\nonumber \\
N_{12}(t) = N_{12}(0) e^{-qt}\nonumber\\
N_{21}(t) = N_{21}(0) e^{-pt}\nonumber\\
N_{22}(t) = N_{22}(0) + N_{12}(0)\left(1 - e^{-qt}\right) \, .
\end{eqnarray}

The consensus function is $C(t) = (1/N)  | N_{11}(t) + N_{12}(t) - N_{21}(t) - N_{22}(t) | $. At zero temperature ($p=q=1$), when the initial beliefs have a 50-50 split and the internal fields are also 50-50 (but independently of the initial beliefs), then $N_{ij}(0)=.25$ independently of $i$ and $j$, and $C(t) = 0$ for all time. Regardless of the distribution of initial beliefs, if the internal fields have the same distribution, the consensus function will be constant in time. Specifically, suppose that the probability of an agent having an initial belief of $+1$ is $r$ and having an initial belief of $-1$ is $1-r$. Suppose also that the probability of an agent having $h_i=+1$ is again $r$, with a probability $1-r$ of having $h_i=-1$. As always, initial beliefs and internal fields are assigned to the agents independently. This gives $N_{11}(0)=r^2$, $N_{12}(0)=N_{21}(0)=r(1-r)$, and $N_{22}(0)=(1-r)^2$. Inserting these values into Eq.~(\ref{eq:solution}) yields $C(t)=2r-1$, independent of time.

\begin{figure}[!h]
\centering 
\includegraphics[scale=0.75]{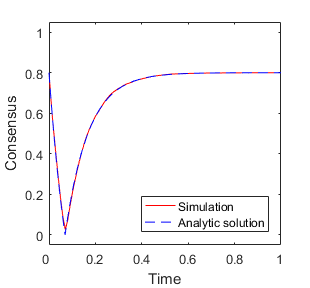}
\caption{Time evolution of the consensus function when the initial distribution of beliefs is 10-90 but internal preference fields are 90-10. The dashed blue line is the analytic solution, using Eqs.~(\ref{eq:solution}) and initial conditions described in the text, and the solid red line is the result of numerical simulations.}
\label{fig:alpha0} 
\end{figure}

The interesting case is when the two differ. For example, suppose that the initial belief distribution has a 10-90 split, but the internal field distribution is 90-10. This case corresponds to an initially rare belief, but one that suits the internal preferences of a majority of the population. For this particular case, $N_{11}(0) = .09 N$, $N_{12}(0) = .01 N$, $N_{21} (0) = .81 N$, and $N_{22} (0) = .09 N$. In this case, the consensus drops from an initial value of 0.8 to 0 at $t=\ln 2$ as agents change their initial beliefs, and then rises again to 0.8.  The resulting consensus function appears in Fig.~\ref{fig:alpha0}.

\section{Proof that majority rule leads to full consensus on the complete graph}
\label{subsec:majfull}

Below we present an informal argument (but one which is easily made rigorous) that majority rule on a complete graph will always achieve full consensus at zero temperature, given any initial distribution and $N$ large. We first discuss the case where {\it all\/} neighbors contribute to the determination of the majority, and then to the case of three randomly chosen neighbors, as used in our numerical simulations.

Consider first any distribution other than 50-50, and where all spins (except the one chosen for updating) contribute to a determination of the majority.   Suppose, for example, there is an initial probability of~0.6 assigned to each agent of being in the $+1$ state, and a probability~0.4 of being in the  $-1$ state. Then the initial state will have a ratio of 3:2 of $+1$~spins to $-1$~spins, with fluctuations from that ratio of order $\sqrt{N}$. So for $N$ already greater than 20, the probability of more spins being $-1$ than $+1$ in the initial state is already extremely small.

Now we start the dynamics. No matter which agent is picked, the majority of her neighbors will be $+1$. So whatever her initial state, she has the state $+1$ afterward. But this increases the dominance of $+1$~spins, so the same argument continues to apply for each successive agent when {\it her\/} Poisson clock rings. Therefore the final state must be $+1$ with probability rapidly approaching~1 as $N\to\infty$. (This result is consistent with the famous Condorcet Jury Theorem~\cite{Condorcet} in political science, where this behavior is typically modeled by a cumulative binomial distribution.)

What about when the initial distribution is 50-50? Here, one can also prove that the system achieves complete consensus in any single run --- but half the time the final state 
is uniformly~$+1$ and half the time it's uniformly~$-1$. The reasoning is similar to that above, with the added ingredient that the initial state will have $\sqrt{N}$ fluctuations that
lead to a small excess of either $+1$ or $-1$~spins with equal probability. But this excess is sufficient that every dynamical realization starting from the initial state will then lead to the uniform final state fully predetermined by the initial excess of plus or minus spins. The {\it only\/} situation where the dynamical realization determines whether the final state is all-plus or all-minus is that where the initial spin configuration has exactly half $+1$ spins and half $-1$ spins, plus or minus a single spin. In this case, the initial dynamics will quickly lead to a small excess of either $+1$ or $-1$ spins with equal probability, again leading to full consensus, but with the final state now determined by the dynamics rather than the initial condition. In any case, the probability of occurrence of such an initial state goes to zero as $N^{-1/2}$. 

This set of arguments, using the usual Curie-Weiss Hamiltonian rather than majority rule (both of which result in identical dynamics at zero temperature), leads to a conclusion significantly stronger than that of the system simply achieving full consensus (i.e., a uniform final state). The stronger conclusion is that the final state is {\it completely\/} predetermined by the initial spin configuration (i.e., all dynamical realizations lead to the same final state) with probability approaching one as the number of spins $N\to\infty$~\cite{YGMNS17}.

\begin{figure}[!h]
\centering 
\includegraphics[scale=0.75]{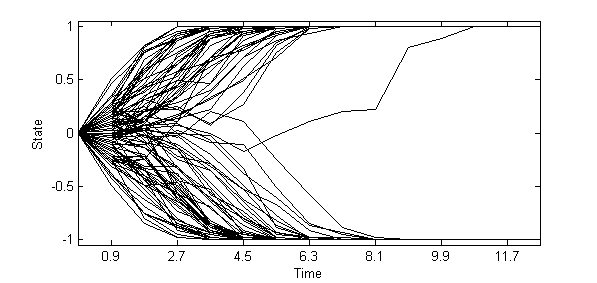}
\caption{Individual runs using majority rule dynamics on the complete graph ($N=90$) with an initial distribution of 50-50.} 
\label{fig:5050} 
\end{figure}

Simulations demonstrating this behavior for an initial 50-50 distribution on the complete graph are shown in Fig.~\ref{fig:5050}. As predicted, full consensus is always achieved, with half the outcomes in the uniform~$+1$ state and half in the uniform~$-1$ state.

For the case when only three agents are chosen randomly to determine the majority, the arguments become more detailed, but the conclusion is the same. An easy way to see why this should be so is to consider the dynamical stable states of the system. A dynamically stable state is a fixed point of the dynamics: once in that spin configuration, {\it no\/} realization of the dynamics can lead to a different state. For the complete graph, the only dynamically stable states under majority rule (and also voter rule, but not expert rule --- cf. Supplement 2~\cite{supp}) are all agents adopting the state~$+1$ or all adopting~$-1$. The main difference between majority rule using {\it all\/} neighbors vs.~three randomly chosen neighbors is that the former case leads to a final state completely determined by the initial spin configuration, while in the latter case the dynamical realization also plays a role in determining the final state. (For a general discussion of the competing roles of initial conditions and dynamical realizations in determining final states, see~\cite{YGMNS17,YMNS13}.)

Importantly, these arguments and results do not hold for locally connected networks: for example, in the weak community structure, {\it any\/} community (i.e., social circle) will be dynamically stable if all its agents share the same state, regardless of what's happening outside the community. Therefore, if $N_c$ is the number of communities in the entire system, there are already $2^{N_c}$ such states, which serves as a lower bound for the number of metastable (i.e., nonuniform but dynamically stable) states under majority rule.

Of course, when the initial condition is extremely skewed, high or full consensus will also be achieved for most other types of networks.  

\clearpage

\section{Additional figures for Section IV}
\label{subsec:empiricalfigs}

\begin{figure}[!h]
\centering 
\includegraphics[scale=0.6, trim= 0 50 0 50, clip]{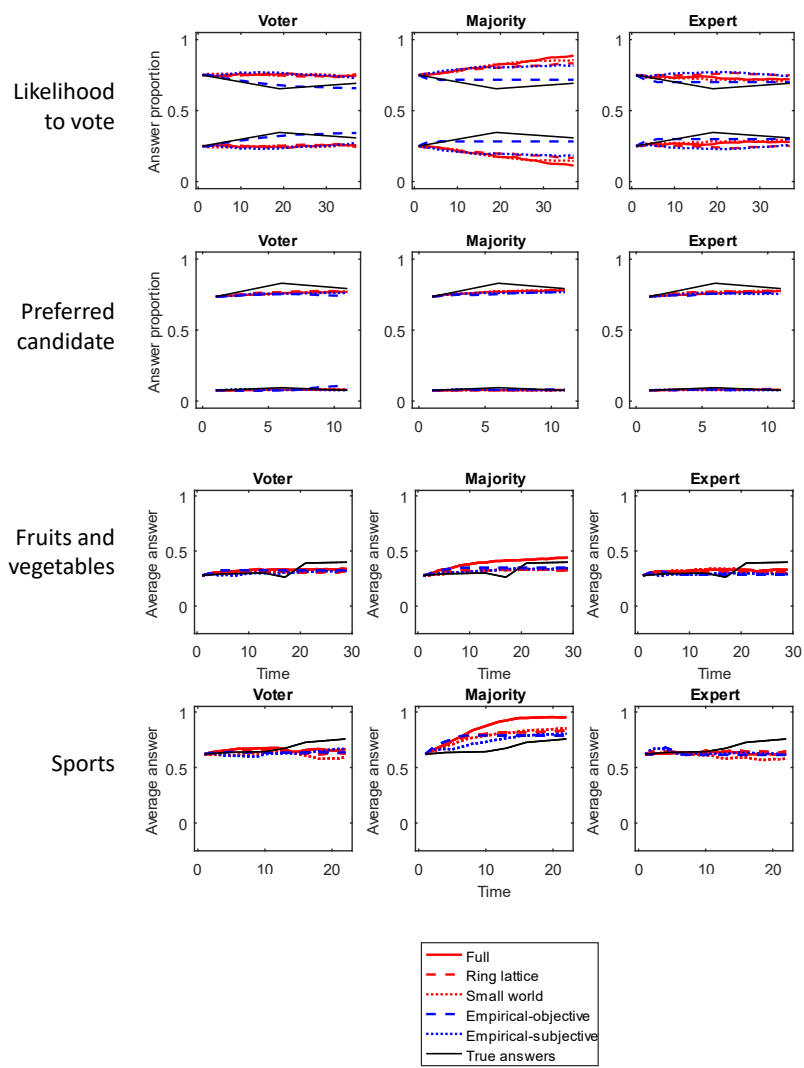}
\caption{Empirically observed and modeled proportion of different responses for  questions in the MIT study.}
\label{fig:MIT_answers} 
\end{figure}

\begin{figure}[!h]
\centering 
\includegraphics[scale=0.7, trim= 0 320 30 0, clip]{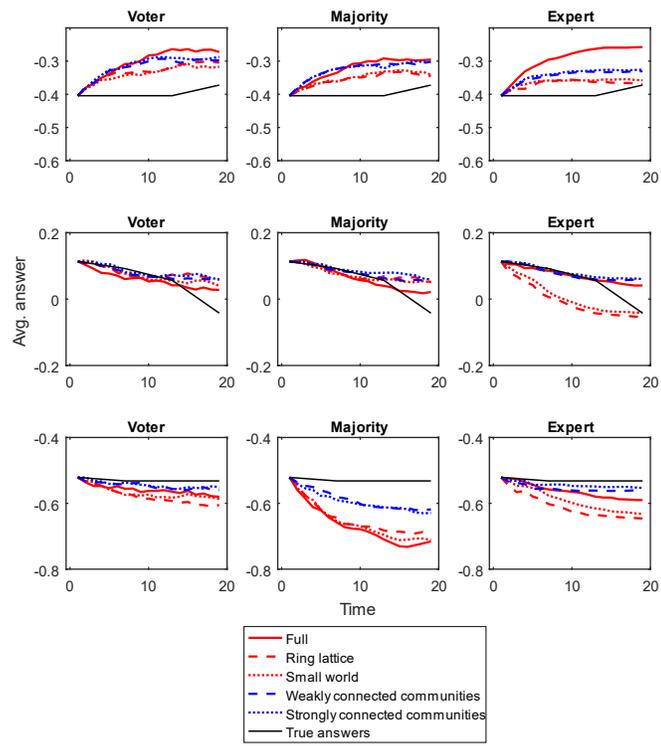}
\caption{Empirically observed and modeled proportion of different responses in the Mechanical Turk study.}
\label{fig:Turk_answers} 
\end{figure}

\clearpage

\end{document}